\documentclass[12pt,letterpaper]{article}
\usepackage{amsmath,amssymb,pgf,pgfarrows,pgfnodes,float,appendix, hyperref}
\usepackage{graphicx}
\usepackage{subfigure}
\usepackage[margin=0.9in]{geometry}

\title{{\rm\footnotesize \qquad \qquad \qquad \qquad \qquad \ \qquad \qquad \qquad \ \ \ \ \ \                  RUNHETC-2018-27}\vskip.5in Broken Scale Invariance Ward Identities for Plasmon Correlators of the Homogeneous Electron Gas}
\author{Tom Banks\\
Department of Physics and NHETC\\
Rutgers University, Piscataway, NJ 08854\\
E-mail: \href{mailto:tibanks@ucsc.edu}{tibanks@ucsc.edu}}

\date{Sept 25, 2018}
\begin{document}
\maketitle

\begin{abstract}  
We derive exact  equations  for a broken  scale invariance  of the  homogeneous  electron gas HEG, and show that they lead to a closed non-linear integral  equation  for the density- density  correlation  function  when evaluated to leading order in the 1/N  expansion.  More generally,  the  identity leads  to  a sequence  of more  refined  systems  of equations,  which close on a finite number  of one plasmon  irreducible  (1PLI)  correlation  functions.
\end{abstract}

\section{Introduction}

The homogeneous electron gas (HEG)  lies at  the foundation  of atomic and condensed matter physics.  The  Schwinger Effective Action  of the  plasmon  field\cite{effact} for this  model is a time  dependent  generalization  of the  Density  Functional of Hohenberg and Kohn\cite{dft}.  If one knew it, it  could be used to compute  the  ground  state  energy of the  gas in an arbitrary background potential.   Choosing that  potential  to be that  of a collection of static  nuclei, and  adding  the Coulomb  repulsion  of the  nuclei,  one has  reduced  the  Born-Oppenheimer approximation  to atoms,  molecules and  solids to  a classical variational  problem.   The  HEG  is not  just  a toy problem.

\section{The Effective Action and Broken Scale Invariance}

In \cite{dftfft}, the author  proposed a computation of the effective action in a systematic  1/N expansion. We will work in units where distances are measured in Bohr radii and energies are measured in Rydbergs.  All coordinates,  fields and parameters in this paper are dimensionless. The classical imaginary time action for the HEG is

\begin{equation} S =  \int d^3x\ dt\ [\frac{N}{4}\phi \nabla^2 \phi + \Psi_a^{\dagger} (\partial_t - \nabla^2 + \mu + i\phi) \Psi_a] . \end{equation}

$a$ is the electron spin component, which we will allow to take values from $1$ to $N$.  The experimental system has $N = 2$.
 This action is believed to be ultraviolet  finite, except for normal ordering, which is equivalent to an additive  shift in $\mu$. We will always compute things in terms of the renormalized chemical potential.    In using this  action  to  compute  the  energy density  of the  model as a functional integral,  we must  divide through  by the  Gaussian  functional  integral  over the  plasmon  field
$\phi$, in order to get the proper Hubbard-Stratonovich transformation.  This has no effect on correlation  functions, which are  ratios of two functional  integrals.
The large N  expansion is generated  by enlarging the number  of fermion spin components from 2 to N , and  multiplying  the  purely  plasmon  term  in the  action  by N/2, as we have done.   For  large N , the quantum fluctuations  of the plasmon field are small and the leading term in the expansion comes from integrating out the fermions and solving the classical equations for $\phi$. In \cite{dftfft} I argued that  this approximation yielded a first order phase transition between a homogeneous gas and a Wigner crystal.   The spin polarized gas phase expected  in three  dimensions from Quantum Monte Carlo simulations  does not make an appearance:  there is no spontaneous  breakdown of $SU (N )$.
In this paper  we will exhibit  a broken scale invariance  Ward  identity,  which may turn  out to be useful in the search for second order phase transitions in this  model.  WeÕll see that  in the large $N$  approximation it leads to an infinite sequence of more and more refined closed systems of integro-differential  equations,  each of which involves only a finite number of correlation functions.  It may come as a surprise that  our model has any remnant of scale invariance,  since everything  is written  in terms  of dimensionless variables.   Nonetheless,  it  is easy to  exhibit the broken symmetry. 

This is most easily done by redefining the fields and coordinates according to
\begin{equation} \phi ({\bf x}, t) = \frac{\mu}{\sqrt{N}} \chi ({\bf y}, s) , \end{equation}
\begin{equation} \Psi_a ({\bf x} , t) = \mu^{3/2} \psi_a ({\bf y} , s) , \end{equation}
\begin{equation} t = a^2 s, \end{equation}
\begin{equation} {\bf x} = a {\bf y} . \end{equation}
The effective action for the ratio of functional integrals, which generates correlation functions of $\chi$ is
\begin{equation} S_{eff} = \frac{\mu^{1/2}}{4} \int d^3 y\ ds [\chi \nabla_y^2 \chi ] + N {\rm ln\ det}\ (\partial_s - \nabla_y^2/2 + 1 + i \frac{\chi}{\sqrt{N}} ) . \end{equation}
We can see immediately that the ordinary perturbation series is a large $\mu$ expansion and the large $N$ expansion is a semi-classical expansion for $\chi$.   In the Wigner crystal phase, we would shift by an order $1$ classical periodic solution of the equations of motion for $\phi$ before writing this form for $S_{eff}$.  The resulting expansion is more complicated and we will not deal with it in this paper.

Now note that  for a static  external  source for $\phi$, ${\cal J} (t, {\bf x}) = i\sqrt{N} {J} ({\bf x})$, $ W [{\cal J} ] = T E[V ({\bf x})]$, where E  is the ground state  energy of the interacting electron gas in the presence of an external potential $ \nabla^2 V = \cal{J}$,  and $T$  is the  length  of the imaginary  time  interval.   Thus,  knowledge of the  effective action,  like that  of the  Density  Functional, reduces  the  Born-Oppenheimer  approximation to  a classical variational  exercise. The effective action approach  to electron dynamics has been championed by [1]. It  is similar in spirit  to Dynamical  Mean Field Theory  for lattice  models, in that  the effective action contains  information  about  the excitation  spectrum  of the model, which is not captured  by the Density Functional. The connection between $\Gamma [\chi]$, the Legendre transform of $W[{J}]$, and the density  functional was explained in \cite{dftfft}. The plasmon field is connected  to the fermion charge density  by GaussÕ law
\begin{equation} \frac{N}{2} \nabla^2 \phi = - i \Psi_a \Psi_a ,\end{equation} 
so its correlation functions are simply related to those of the electron charge density.   

It is possible to derive the equations for broken scale invariance by functional manipulations
using these definitions, but  in the interests  of clarity  I will present the derivation  in the next section in the language of Feynman  diagrams of the 1/N expansion for the correlation functions of the plasmon field.
It  is a  well known  consequence  of the  algebraic  properties  of Legendre  transforms  that
$W_n (p_1, . . . p_n )$ is a sum  of all tree  diagrams  with  vertices  made  from one plasmon irreducible correlators $\Gamma_m$  with  $m \leq n$,  and
limbs made from the full propagator $W_2 (p)$.  Written in terms of $\Gamma_m $, the scaling Ward identity
is a highly non-linear  equation  relating  1PLI  vertices  to  those  with  more legs.   In this,  itÕs
similar to the Schwinger-Dyson equations.   WeÕll see however that  this  hierarchy  truncates in an interesting  way in the $1/N$  expansion.

\section{$1/N $ Expansion of the  Scaling Ward Identity}

It  is clear  that   for large  $N$ ,  $\phi$ is a  semi-classical variable.    In  the  gas  phase  the  classical configuration  around  which we expand  is $\phi = 0.$   Note  that  the  transformation $\phi (x, t) \rightarrow
 \phi (x, t) + \lambda (t)$  is a gauge transformation. That  is, the zero wave number  mode of $\phi$ decouples from the  fermion determinant.  To do this  carefully,  we should  work on a spatial  torus  and simply discard  the  discrete  zero mode.   We will simply make  sure that  our calculations  are
 consistent with gauge invariance.  The k point term in the action, $S_{eff} [\chi]$, which is the leading approximation to the functional $\Gamma [\chi_c]$ scales like $N^{(2-k)/2}$.  The $1/N$ expansion of $\Gamma_k (q)$ in the homogeneous phase consists of all $k$ point Feynman diagrams, with vertices given by the coefficients in the expansion of $S_{eff}$, which are single loop fermion diagrams with $k$ legs  and propagator

\begin{equation} D(q) =[ \frac{\mu^{1/2}}{4} {\bf q}^2 + S({\bf q}, \omega)]^{-1} , \end{equation} where
$S$ is the familiar Lindhard  function, with chemical potential  set equal to 1. The dependence of $\Gamma_k (q)$ on $\mu$ for fixed $q_i$  comes only from differentiating  the internal  propagators  as shown in the Figure.

\begin{figure}[h]
\begin{center}
\includegraphics[scale=0.5]{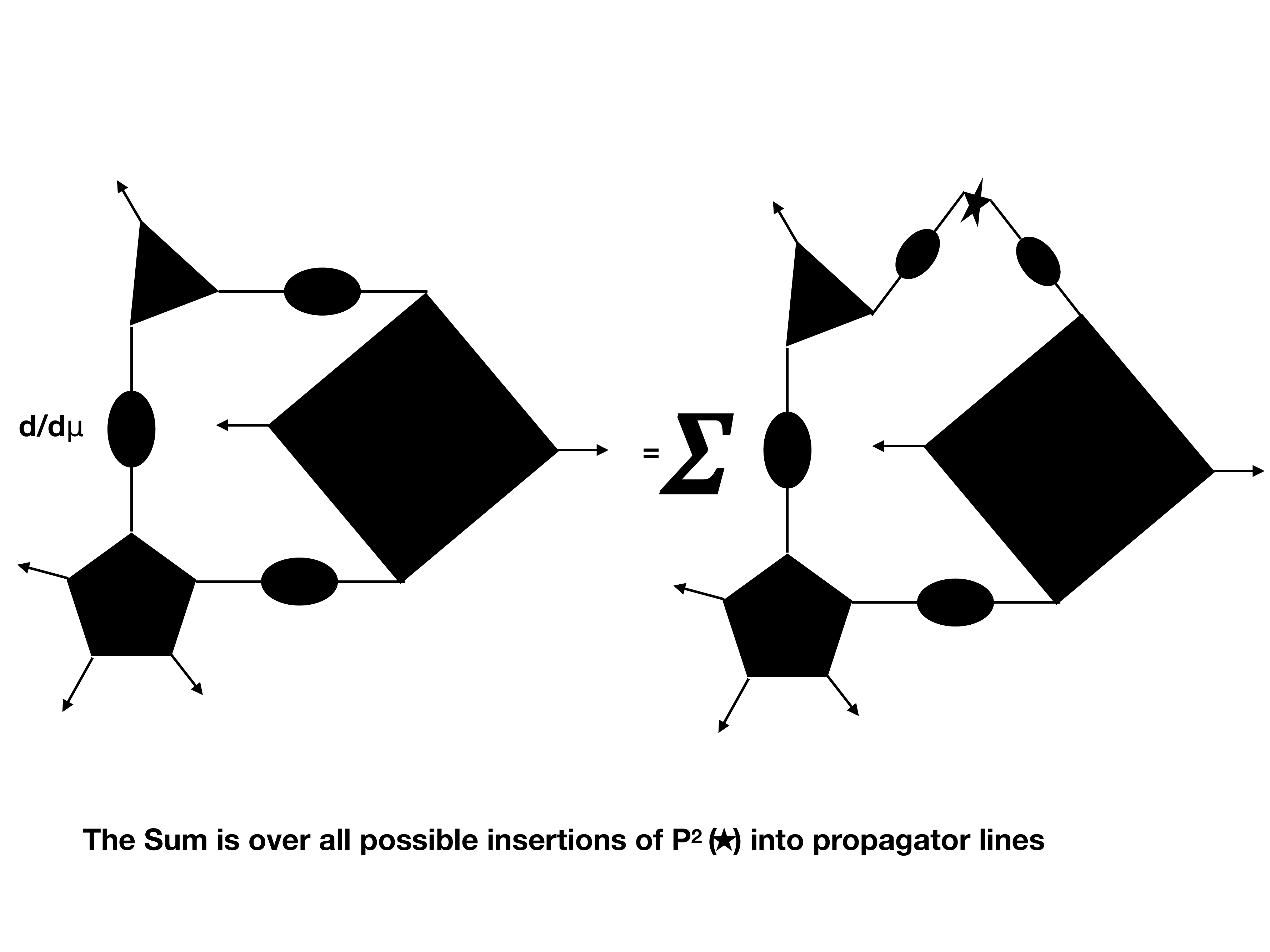}
\caption{Graphical derivation of the scaling relation.}
\label{z}
\end{center}
\end{figure}

Differentiating a propagator gives
 
\begin{equation} \mu\partial_{\mu} D(q) = -  \frac{\mu^{1/2}}{8} {\bf q}^2 D^2 (q) . \end{equation}  
This differentiated  propagator is integrated against  a k + 2 point  function,  with momenta  $q,
-q,q_1\ldots q_k .$  This function is connected,  but  not necessarily 1PLI, because the derivative
has broken open one internal  propagator. In fact, every diagram for the connected k + 2 point
function  contributes  to the  scaling derivative  of the  1P LI  two  point  function.   However, one must be careful about  the propagators  on the external  legs of the k + 2 point function.  On the k legs of the  original 1PLI vertex,  there  are no propagators, so, in the  popular  jargon,  these
legs are truncated.  On the  integrated legs, with  momentum  $q$ and  $- q$, we resum the  Dyson series and  get $W_2 (q)$, the inverse $\mu^{1/2} {\bf q}^2 /4 + \Pi ({\bf q}, \omega)$ where $\Pi$ is the  full polarization  function,  summed  to all orders in the  $1/N $ expansion.  The  easiest way to see that  the  combinatorics  works out  is to go through  the functional  derivation  in the appendix.  Defining $z = - {\rm ln}\mu$ we get an exact equation

\begin{equation} \partial_z \Gamma_k (q_1 \ldots q_k ) = - \frac{\mu^{1/2}}{8} \int \frac{d^4 q}{(2\pi)^4} {\bf q}^2 W_2 (q) W_2 (- q) W_{k + 2}^T (q_1 \ldots q_k, q , - q) + \delta_{k2} \frac{\mu^{1/2}}{8} {\bf p_1}^2 \delta^4 (p_1 + p_2). \end{equation}

 In this equation,  the connected correlator $W_{k + 2}$ has all of its legs truncated because we have exhibited explicitly the external propagators on the integrated legs.
Like the Schwinger-Dyson equations in models with cubic and quartic interactions,  this equation relates 1PLI correlators  with k points to those with a larger number  of points.  Since the action  for the plasmon  field contains  vertices of all orders in $\chi$, the SD equations following from the effective action are much more complicated.  In principle the scaling equation
involves $\Gamma_{k+1}$ and $\Gamma_{k+2}$ , and so does not truncate. Recall however, that $\Gamma_k  \sim N^{1 - k/2}$ , plus higher
orders in $1/N$ . This suggests an approximation scheme in which we replace $\Gamma_{k+1}$ and $\Gamma_{k+2}$ by
their leading large $N$ approximation and get a closed system of scaling equations involving only
$\Gamma_m$ with $m \leq k$.  The simplest such approximation is a closed equation  for $\Gamma_2$ :

\begin{equation} \partial_z \Gamma_2 (p) =\frac{ e^{-z/2}}{4N}\int \frac{d^4 q}{(2\pi)^4} {\bf q}^2 \Gamma_2 (q)^{-1} \Gamma_2 (- q)^{-1} [S_4 (q, - q, p, - p)  +\end{equation}\begin{equation} S_3 (q, - q, 0)S_3 (p, - p, 0)\Gamma_2 (0)^{-1} + S_3 (q,p, - q - p) S_3 (-q, - p, p + q)) \Gamma_2 (p + q)^{-1} \end{equation}\begin{equation} + S_3 (q, - p, p - q) S_3 (-q, p, q - p) \Gamma_2 (p - q)^{-1} . \end{equation}  In this equation $S_{3,4}$ are the one fermion loop vertices of 
$S_{eff}$ stripped of their powers of $N$.
The  action  for the  Plasmon  field is non-polynomial so its S-D equations  are extremely  complicated.  The scaling equation  is more analogous  to the coupled boson-fermion  S-D equations,  but  purely  in terms  of bosonic correlators.  This equation is supplemented  by the large µ boundary  condition that $\Gamma_2 \rightarrow \frac{{\mu}^{1/2}}{4} {\bf p}^2 + S(q) $. 
The solution of the scaling equation  with these boundary  conditions  is a complicated function of $N$  and $z$ and might capture  some of the phase structure of the model at finite N .
It would be particularly interesting  to explore the question of whether the solution can have zero frequency singularities  as the spatial  momentum  vanishes, since these would indicate  the existence of a gapless plasmon excitation,  and would be a likely sign of a quantum critical point.
 
The  equations  for $\Gamma_{3,4}$ involve
 $\Gamma_{5,6}$  and  one can obtain  a more refined approximation by
 making the substitution $\Gamma_{5,6} \rightarrow S_{5,6}$ . This gives a highly non-linear set of coupled equations for $\Gamma_{2,3,4}$.  One would imagine that  this  second set of equations  captures  much of the low energy
dynamics of the HEG. High point correlation  functions contain  only rather  complicated  multi-
plasmon  interactions, and  probably  do not  contribute much to the  coarse grained  properties studied  in experiment.  Approximating  them by their leading order large N , behavior which is also dominant in the high density limit, seems quite innocuous.
Once one has chosen one of these  approximation schemes, one computes  the  plasmon  effective action  by using the  solutions  to  compute  the  first few terms  in the $\phi$ expansion  and approximating the rest by their leading large $N$  behavior.  One then  has a classical variational problem  to  solve in order  to  find the  Born-Oppenheimer potential.    It  seems clear that  any serious attack  on these equations  will have to be numerical,  but perhaps  analytical  insight can be gained by looking for solutions with some kind of scaling behavior.

\section{Conclusions}

We have proposed a sequence of approximation schemes, each a refinement of the $1/N$  expansion for the  HEG. The  simplest  of these,  which is a closed equation  for the  two  point  function  of the plasmon field, simply related  to the density density correlation  function, deserves the most attention. One should try to analyze the possibility of a gapless plasmon and or scale invariant behavior.  More straightforwardly, one can try to solve the equation  numerically.  The unknown function  depends  on three  variables,  and  is a simple evolution  equation  in one of them.   It remains to be seen how difficult a numerical challenge this will be.
ItÕs also important to find the corresponding  set of equations  in the Wigner crystal  phase of the  model.  The  background  classical solution  provides an additional  breaking  of the  scale symmetry weÕve used, and various contributions to the scaling equations that  vanished because of exact  moment  conservation  will now have  umklapp  contributions.  The  equations  will be much  more complex,  but  might  be revealing.   We could also try  to  generalize our  analysis to phases of the system with $SU (N )$ breaking,  since quantum  Monte Carlo Methods  seem to indicate  that  such phases occur for $N  = 2$.  ThereÕs a very general argument that  breaking to
$SU (M ) \times SU (N - M )$ is only possible at large $N$  when $M$  is of order $1$. The free energy of the
model scales like $N$  at large $N$ , so the system cannot  have more than  $o(N )$ Goldstone  bosons.
Large N  semi-classical analysis does not show any instability  in these Goldstone  directions.  If the approximate  scaling identities  do detect  the broken symmetry  phase, that  would be strong evidence for their utility.
 
\section{Appendix: Functional Derivation of the  Scaling Equations}

The action for the large N  fluctuation  field $\chi$ is
\begin{equation} S_{eff} = \frac{\mu^{1/2}}{4} \int d^3 y\ ds [\chi \nabla_y^2 \chi  ] + N {\rm ln\ det}\ (\partial_s - \nabla_y^2/2 + 1 + i \frac{\chi}{\sqrt{N}} ) + \int d^3 y\ ds J \chi. \end{equation}
The  derivation  of the  scaling equations  is now straightforward.  The generating functional $Z[J]$ is the ratio of the functional integral with this action, and that with $J = 0$.  Thus ($\mu = e^{-z}$)
\begin{equation} Z^{-1} \partial_z Z = \frac{\mu^{1/2}}{8} \int d^3 y\ ds  \nabla_y^2 [Z^{-1}\frac{\delta^2 Z}{\delta J(y) \delta J(y)}  -  \frac{\delta^2 Z}{\delta J(y) \delta J(y)}_{J = 0}] . \end{equation} In this equation $y$ is the four vector $(s, {\bf y})$ and the action of $\nabla_y^2$ on the coincident point two point function is defined by taking a limit of its action for non-coincident points. These terms are not integrals of total derivatives, and are better understood in momentum space.  The second term in square brackets is independent of $J$, and will disappear when we look at equations for connected correlation functions.

We do this by writing $Z = e^W$ and differentiating with respect to $J$.
\begin{equation}  \partial_z W = \frac{\mu^{1/2}}{8} \int d^3 y\ ds  \nabla_y^2 [\frac{\delta^2 W}{\delta J(y) \delta J(y)} + \frac{\delta W}{\delta J(y)} \frac{\delta W}{\delta J(y)} ].\end{equation}

Knowledgeable  readers  will notice  the  structural relation  between  these  equations  and  the Wegner-Wilson-Polchinski  exact renormalization group equations.  If we rewrite the equations for the generating functional of 1PLI correlators, the second term in brackets only contributes to the equation for the two point function.  In terms of the 1PLI generating functional our equation reads
\begin{equation} \partial_z[\Gamma[\chi_c] - \int d^4 q\ \chi_c (q) \frac{\delta \Gamma}{\delta \chi_c (q)}] = \int \frac{d^4 q}{(2\pi)^4} {\bf q}^2 [ \frac{\delta^2 W}{\delta \chi_c (q) \delta \chi_c (-q)} + \chi_c (q) \chi_c (-q)] . \end{equation}  we havewritten this in momentum space because it clarifies the action of $\nabla^2$ on the right hand side and because it simplifies our integral equations. If we take $k$ derivatives with respect to $\chi_c$ on the left hand side, we get $(1 - k) $ times the k-point 1PLI vertex.  Recalling that $\frac{\delta J(p)}{\delta \chi_c (q)}  = W_2^{-1} (p,q) $ (where the inverse is meant in the sense of integral operators) we see that we can write the equations as
\begin{equation} (1 - k) \partial_z \Gamma_k (p_1, \ldots p_k) = \int \frac{d^4 q}{(2\pi)^4} {\bf q}^2 [ W_2 (q) W_2 (-q) W_k^T (q,-q, p_1, \ldots p_k)] + \delta_{k2} \frac{e^{-z/2}}{8} {\bf p_1}^2 \delta^4 (p_1 + p_2). \end{equation}  
$W_k^T$ is the connected $k-$ point function with all of its legs truncated.  It has a tree diagram expansion involving all of the $\Gamma_l$ with $l\leq k$.

Like Schwinger-Dyson equations in theories with cubic and quartic interactions, this hierarchy relates the infinite set of 1PLI correlators to each other.  However, we can truncate it at any $k$ by approximating $\Gamma_{k + 1}$ and $\Gamma_{k + 2}$ by $N^{1 - k/2} S_{k + 1}$ and $N^{1/2 - k/2} S_{k + 2} $, and obtain a closed set of equations.  The equation in the text is what we obtain for the $k = 2$ truncation.

\vfill\eject
\vskip.3in
\begin{center}
{\bf Acknowledgments }\\
The work of T.Banks is {\bf\it NOT} supported by the Department of Energy, the NSF, the Simons Foundation, the Templeton Foundation or FQXI. I'd like to thank Bingnan Zhang for proofreading the manuscript and correcting a number of errors.\end{center}


\begin{thebibliography}{99}
\bibitem{effact}Yi-Kuo Yu, ``Derivation of the Density Functional via Effective Action" arXiv:09100670v3[cond-matt.other] . 
\bibitem{dft} Hohenberg, P.; Kohn, W. (1964). "Inhomogeneous Electron Gas". Physical Review. 136 (3B): B864. Bibcode:1964PhRv..136..864H. doi:10.1103/PhysRev.136.B864.
\bibitem{dftfft}  T.~Banks, ``Density Functional Theory for Field Theorists I,''
  arXiv:1503.02925 [cond-mat.mtrl-sci].

  
  \end{thebibliography}
\end{document}